\documentstyle[12pt,a4]{article}

\begin{document}

\newcommand{\ga}{\gamma}
\newcommand{\C}{{\bf C}}
\newcommand{\R}{{\bf R}}
\newcommand{\N}{{\bf N}}
\newcommand{\Z}{{\bf Z}}
\newcommand{\al}{\alpha}
\newcommand{\be}{\beta}
\newcommand{\Q}{{\bf Q}}
\newcommand{\pf} { {\rm {\bf Proof} } }
\newcommand{\EOP} { \hfill $\Box$ }
\newcommand{\pa} { \partial }

\newtheorem{thm}{Theorem}
\newtheorem{lemma}[thm]{Lemma}

\title{On rational definite summation\thanks{The paper
got partial financial support from a RFBR grant 04-01-00130}}
\author{
 Sergey P.\ Tsarev\\
 {Department of Mathematics, } \\
 {Krasnoyarsk State Pedagogical University} \\
 {Lebedevoi 89, 660049 Krasnoyarsk, Russia}\\
 {\tt tsarev@newmail.ru}
 }

\date{}

 \maketitle
\begin{abstract}
We  present a partial proof of  van~Hoeij-Abramov conjecture about
the algorithmic possibility of computation of finite sums of
rational functions. The theoretical results proved in this paper
provide an algorithm for computation of a large class of sums
(\ref{s1}).
\end{abstract}

\section{Introduction} \label{SECT:intro}

The problem of finding sums
\begin{equation}\label{s1}
  S(n) = \sum_{k=0}^{n-1}R(k,n)
\end{equation}
algorithmically in ``closed form" for various terms $R(k,n)$ was
intensively studied in the last decades (see \cite{A}, \cite{A02},
\cite{AL00}, \cite{PWZ} and references therein). For
hypergeometric $R(k,n)$ which do not depend on $n$ (the so-called
"indefinite summation problem") it was completely solved in
\cite{G}; some further research was done in order to make the
Gosper's algorithm more efficient.  When the terms $R(k,n)$ depend on $n$
explicitely (the so-called "definite summation problem")
 S.A.Abramov and  H.Q.~Le in
\cite{A02}, \cite{AL00} gave necessary and sufficient conditions
for termination of the well-known Zeilberger algorithm for
rational $R(k,n)$ and general hypergeometric $R(k,n)$ thus giving
the evidence of incompleteness of the existent algorithmic
methods in this general case. Recently M.~van~Hoeij \cite{MvH},
generalizing previous observations by S.A.Abramov, proposed a conjecture
which (if proved) would provide a complete and relatively simple
algorithm for finding sums (\ref{s1}) for rational
$R(k,n)=P(k,n)/Q(k,n) \in \Q[k,n]$ if the sum $S(n)$ is again a
rational function of $n$. In this paper we give a partial proof of
this conjecture, namely we prove that for denominators $Q(k,n)$
which factor in $\overline{\Q(n)}[k]$ (i.e.\ over the field of algebraic
functions of $n$) as
\begin{equation}\label{Q}
  Q(k,n) = \prod_{s=1}^p (k-\al_s(n))^{d_s}
\end{equation}
and the {\em algebraic} functions $\al_s(n)$ have asymptotics
$\al_i(n)\sim c_i n^{\varepsilon_i}$ with $\varepsilon_i < 1$ for
$n\rightarrow\infty$,  one can algorithmically find out if the sum
$S(n)$ is rational and if so find it using some simple
rearrangements of terms in (\ref{s1}) proposed by  S.A.~Abramov
and M.~van~Hoeij. Such $Q(k,n)$ we will call hereafter ``slow
denominators".

The paper is organized as follows. In the next section we explain
how one can transform the sum (\ref{s1}) in the general case into
a finite expression in terms of the polygamma functions and
algebraic functions. In section~\ref{SECT:polygamma} we expose the
transformations of the sum (\ref{s1}) proposed by S.A.~Abramov and
M.~van~Hoeij and explain their link to the well-known identities
for the polygamma functions. In that section we also formulate
two questions about the possibility of nontrivial new identities
for the polygamma functions which are crucial for the problem of
definite rational summation. Sections~\ref{SECT:C}
and~\ref{SECT:N} give the main theoretical results of this paper:
the first question is answered completely, for the second we do this
only for ``slow denominators".
In the concluding section we discuss some algorithmic questions.

\section{Preliminaries} \label{SECT:pre}

We always implicitely assume that all terms in (\ref{s1}) are
nonsingular: $Q(k,n) \neq 0$ for $0 \leq k \leq n-1$, $k \in
\Z$ at least for $n$ large enough.
 After decomposition of $R(k,n)=P(k,n)/Q(k,n)$ into a sum of a polynomial and
elementary
fractions $R = \overline P(k,n) + \sum_s \left(\sum_{t=1}^{d_s}
\frac{\be_{st}(n)}{(k-\al_s(n))^{t}}\right)$ with algebraic
$\al_s(n)$, $\be_{st}(n)$ using (\ref{Q}) we can represent
(\ref{s1}) as a finite sum of a polynomial and of "atomic" sums $\sum_{k=0}^n
\frac{1}{(k-\al_s(n))^{t}}$ with algebraic multipliers
$\be_{st}(n)$. Each atomic sum may be expressed as a difference
\begin{equation}\label{Atom}
  \sum_{k=0}^{n-1} \frac{1}{(k-\al_s(n))^{t}} = \psi^{(t-1)}(n-\al_s(n))
    - \psi^{(t-1)}(-\al_s(n))
\end{equation}
using the standard identity
\begin{equation}\label{psi1}
 \psi^{(s)}(z+1) = \psi^{(s)}(z) + \frac{1}{z^{s+1}},
\end{equation}
where $\psi^{(s)}(z)$ is the normalized polygamma function
$$\psi^{(s)}(z)=(-1)^{s} d^{s+1}(\log\Gamma(z))/(s!dz^{s+1})$$
(cf. \cite{AS}, \cite{BE}). We also use the standard notation
$\psi(z)=d(\log\Gamma(z))/dz= \Gamma'(z)/\Gamma(z)$ for $s=0$.
 So (\ref{s1}) may be transformed into a finite sum of the form
\begin{equation}\label{s2}
  \sum_i \be_i(n)\psi^{(t_i)}(\gamma_i(n))
\end{equation}
with algebraic $\be_i(n)$, $\gamma_i(n)$.


\section{Rearrangement of terms and  polygamma identities} \label{SECT:polygamma}

Obviously one can transform any atomic sum in (\ref{Atom}) or
directly (\ref{s1}) into an equal sum substituting $k \rightarrow
n-k$:
\begin{equation}\label{n-k}
\sum_{k=0}^{n-1} \frac{1}{(k-\al(n))^{t}} =
 \sum_{k=0}^{n-1} \frac{1}{(n-1-k-\al(n))^{t}} =
 (-1)^t \sum_{k=0}^{n-1} \frac{1}{(k-\overline\al(n))^{t}},
\end{equation}
$\overline\al(n)=n-1-\al(n)$ ("Abramov transformation"). This
corresponds to the known identity
\begin{equation}\label{psi2}
   \psi(1-z) = \psi(z) + \pi \cot(\pi\,z)
\end{equation}
and its derivatives for the higher polygammas $\psi^{(s)}(z)$
after the transformation (\ref{Atom}). Another transformation
proposed by M.~van~Hoeij splits odd and even terms (hereafter we
assume without loss of generality $t=1$ to avoid cumbersome
notations):
\begin{equation}\label{o-e}
\sum_{k=0}^{2n-1} \frac{1}{(k-\al(n))}
 = \sum_{k=0}^{n-1} \frac{1}{(2k-\al(n))}
 + \sum_{k=0}^{n-1} \frac{1}{(2k+1-\al(n))}
\end{equation}
$$ = \frac{1}{2}\sum_{k=0}^{n-1} \frac{1}{(k-\al(n)/2)}
 + \frac{1}{2}\sum_{k=0}^{n-1} \frac{1}{(k+1/2-\al(n)/2)}.
$$
We may choose to split the sum into $m$ parts in a similar way.
 This transformation corresponds to the identity (see \cite{BE})
\begin{equation}\label{psi3}
   \psi(mz) =  \log m +\frac{1}{m}\sum_{r=0}^{m-1} \psi\left(z+\frac{r}{m}\right)
\end{equation}
and its derivatives.
 Alternatively one can split the sum into two consecutive subsums:
\begin{equation}\label{subs}
\sum_{k=0}^{n-1} \frac{1}{(k-\al(n))}
 = \sum_{k=0}^m \frac{1}{(k-\al(n))}
 + \sum_{k=0}^{n-m-2} \frac{1}{(k+m+1-\al(n))}.
\end{equation}
 This transformation trivializes in the language of the polygamma
functions: $\psi(a)-\psi(b)=\left(\psi(a)-\psi(c)\right)+\left(\psi(c)-\psi(b)\right)$.
 Obviously one can also split off or add a fixed number of terms in the
beginning or the end in any sum (\ref{s1}) extracting rational
functions $R(0,n)$, $R(1,n), \ldots $ --- this corresponds to the
basic identity (\ref{psi1}).

Now one can formulate the following natural questions:

{\bf Question 1}. What finite sums of the form (\ref{s2}) with
algebraic $\be_i(n)$, $\gamma_i(n)$ are in fact equal to rational (or algebraic)
functions of $n \in \C$ (that is for arbitrary complex $n$)?

{\bf Question 2}. What finite sums of the form (\ref{s2}) with
algebraic $\be_i(n)$, $\gamma_i(n)$ are in fact equal to rational (or algebraic)
functions of $n \in \N$ (that is only for positive integer $n$)?

We show in the next section that the first question may be easily
answered: the only possible identities of this form valid for {\em
arbitrary complex} $n$ are those obtained from (\ref{psi1}) and
(\ref{psi3}) after substitution of an algebraic function instead of $z$
in the arguments, multiplication with another algebraic function
and addition of several such identities.

The second question is more difficult. As (\ref{psi2}) shows we
can find identities with extra terms which vanish for integer
values of $n$. In section~\ref{SECT:C} we prove that there is no
additional identities in the case of ``slow" $\gamma_i(n)$.

\section{Complex Identities} \label{SECT:C}

\begin{lemma}\label{L1}
 Let $\al(z)$ be an algebraic function such that:

a) $\al(n_i) \in \N$ for infinitely many $n_i \in \N$;
\smallskip

b) $\displaystyle\overline {\lim_{i \rightarrow \infty}}
\frac{i}{n_i}> 0$;
\smallskip

c) $\al(z) \sim \be z$ for $z \rightarrow\infty$ and $\be$ is a
positive real number.

Then $\al(z) = \be z+\gamma$ for constant $\be$, $\gamma$, $\be
\in \Q$, $\gamma \in \Z$.
\end{lemma}
\pf Since $\displaystyle\overline {\lim_{i \rightarrow \infty}}
\frac{i}{n_i}> \frac{1}{N_0}$ for some large  $N_0 \in \N$ 
and we can always assume that  ${n_i}$ is monotonically increasing,
there
exist infinitely many pairs of integers $n_{i_s+1}$, $n_{i_s}$
such that $n_{i_s+1}-n_{i_s} < N_0$.  Then in turn we can choose among them
infinitely many pairs with $n_{i_s+1}-n_{i_s} = N_1$ for some
fixed $N_1 \in \N$. From c) we conclude that $\al'(z) \rightarrow
\be$ for $z \rightarrow\infty$ so for the chosen pairs we have
$|\al(n_{i_s+1})-\al(n_{i_s})| < M_0$. Choose infinite subset of
such pairs with $\al(n_{i_s+1})-\al(n_{i_s}) = M_1$, $M_1 \in \N$.
We have two algebraic functions  which are equal in infinitely
many integer points so they are identically equal:
$\al(z)=\al(z+N_1)-M_1$. Take $z_0 \in \R$ such that $\al(z)$ has
no singularity on the real axis for $z>z_0$ so
$\al(z_0+N_1)=\al(z_0)+M_1$, $\al(z_0+2N_1)=\al(z_0)+2M_1$, \ldots
The algebraic function $\al(z)$ is equal to $M_1(z-z_0)/N_1
+\al(z_0)$ in infinitely many poins so they are equal for all $z$.
\EOP

Suppose we have a finite sum (\ref{s1}) which is a rational
function of $n \in \C$. Then it has a finite number of poles. The
function $\psi(z)$ has simple poles at $z=0$, $z=-1$, $z=-2$,
\ldots\ with residues $-1$; $\psi^{(t)}(z)$ have at the same
points poles of order $t+1$ with  coefficients $-1$ in the Laurent
expansions. In order to have finite number of poles in (\ref{s1})
almost all must cancel. They can not be suppressed by the
algebraic coefficients $\be_i(z)$ which have only finite number of
poles and zeros (on an appropriate Riemannian surface). Consider
an accumulation point of the poles of a given
$\psi^{(t)}(\al_i(z))$, i.e.\ a point $z_0 \in \C$, such that
$\al_i(z)\rightarrow\infty$ for $z\rightarrow z_0$. Take the
highest value of $(t_i)$ in (\ref{s2}). Then there are several
terms with the same $(t)$ and the same accumulation point $z_0$.
Let the corresponding $\gamma_{i_p}(z) \sim c_p
(z-z_0)^{\varepsilon_p}$ for $z\rightarrow z_0$ and
$\varepsilon_0$ be the largest exponent among them. Then obviously
we may consider only the terms with the same value of the exponent
$\varepsilon$ in the asymptotics: the smaller values will give
density  of poles not enough for cancellation with this term.
Suppose $|c_1|$ is the largest absolute value of coefficients in
this asymptotics (probably we have several such $c_i$). Near
$z_0$ make the following change of variables: $w = \al_{i_1}(z)$, so
for $z\rightarrow z_0$, $w\rightarrow \infty$. On the $w$-plane in
the neighbourhoud of infinity we now consider only the terms chosen
above. $\psi^{(t)}(\al_1(z)) = \psi^{(t)}(w)$ has the poles at
$w=0$, $w=-1$, $w=-2$, \ldots, almost all must cancel with the
poles of the finitely many similar terms $\psi^{(t)}(\al_{i_p}(z))
= \psi^{(t)}(\rho_{i_p}(w))$, $\rho_{i_p}(w) \sim \bar c_p w$,
$w\rightarrow \infty$.  We easily conclude that some of
$\rho_{i_p}(w)$ satisfy the hypotheses in Lemma~\ref{L1} so they are
$\rho_{i_p}(w) = \be_p z+\varsigma_p$, $\be_p \in \Q$, $\varsigma_p
\in \Z$. All terms in question can be collected in subgroups with
this property, i.e.\ the arguments in $\psi$ differ only by a transformation
$\ga_{i_p}(n) = \be_p \ga_{i_0}(n)+\varsigma_p$, $\be_p \in \Q$, $\varsigma_p \in \Z$.; 
each subgroup must have separate cancellation of
almost all poles. Considering now the residues of the poles and
using similar asymptotic expansions for the algebraic coefficients
$\be_{i_p}(z)$ we conclude that that the corresponding
coefficients $\be_{i_p}(z)$ in a subgroup form subsubgroups where
they differ only by a constant multiplier and ensure separate
cancellation of almost all poles of this subsubgroup. 
It may be necessary to split some $\be_{i_p}(z)$ into sum of other  $\be_{i}(z)$.
Making
another change of variables $w\rightarrow M w$ for some large
integer $M$ and using (\ref{psi3}) we see that  for each
subsubgroup we obtain a sum of the form $\be(w)\sum_p
k_p\psi^{(t)}(w+n_p)$ with integer $n_p$;  then using (\ref{psi1}),
we reduce it to a rational function of $w$ multiplied by an
algebraic function so it is an algebraic function of the initial
variable $z$ and may be transformed to it algorithmically. In the
process of computation one may use the standard Puiseux
expansions based on the Newton polygon for the algebraic
functions.

Thus we have proved that there are no {\em essentially new}
identities of the form $\sum_i \be_i(n)\psi^{(t_i)}(\gamma_i(n)) =
R(n)$ with rational (and even algebraic) $R(n)$ valid for
arbitrary complex $n$, all such identities may be deduced from
(\ref{psi1}), (\ref{psi3}) and for any sum (\ref{s2}) we may
algorithmically decide if it is a rational (or algebraic)
function.

\section{Integer Identities} \label{SECT:N}

As we have already mentioned the case of identities of the form
\begin{equation}\label{id}
  \sum_i \be_i(n)\psi^{(t_i)}(\gamma_i(n)) = R(n)
\end{equation}
valid only for positive integers $n$ is more difficult. First of
all we restrict our attention to sums of terms
$\be_i(n)(\psi^{(t_i)}(n-\al_i(n))- \psi^{(t_i)}(-\al_i(n)))$
originating from (\ref{Atom}) and we assume that for
$n\rightarrow\infty$, $\al_i(n)\sim c_i n^{\varepsilon_i}$ with
$\varepsilon_i < 1$ (``slow denominators"). 
Consider the poles of such terms in the right
half-plane $Re(n) >0$. The first term $\psi^{(t_i)}(n-\al_i(n))$
obviously has only finitely many of them due to our restriction
on $\al_i(n)$. In order to kill the poles of the second term
$\psi^{(t_i)}(-\al_i(n))$ let us multiply the sum (\ref{s1}) by
$\prod_i \sin^{t_i+1}(\pi \al_i(n))$. Thus we obtain from the
original sum (\ref{s1}) a multivalued function without poles or
ramification points for $Re(n) >n_0$. Take a branch of this
function for $Re(n) >n_0$; after an integer shift we obtain a
function $f(z)$ holomorphic in the right half-plane vanishing at
integers. Using the known asymptotics
 for the polygamma function
in the right half-plane we see that even after multiplication by
the product of sines its growth is bounded as $|f(z)| < A
\exp(\tau |z|^\varepsilon)$ with $\varepsilon < 1$. We can use
{\bf Carlson's theorem} \cite[p.~247]{Boas}:

{\em If $f(z)$ is holomorphic for $Re(z) \geq 0$, $|f(z)| < A
\exp(\tau |z|)$ with $\tau < \pi$, and if $f(n)=0$ for $n=0$,
$n=1$, $n=2$, \ldots , then $f(z)\equiv 0$}.

This reduces the situation to that of section~\ref{SECT:C} so we
conclude the absence of essentially new identities and algorithmic
possibility to decide if a given sum (\ref{s1}) with ``slow"
denominators is rational for sufficiently large integer values of
$n$. 

\section{Concluding remarks} \label{SECT:cr}

Using the transformations (\ref{n-k}) and (\ref{o-e}) and their
analogs for arbitrary $m$ we can make a weaker restriction
$\al_i(n) \sim c_in^{\varepsilon_i}$, $\varepsilon_i \leq 1$, but
at the moment we see no way of treating the general case. Obviously
the algorithm that uses our theoretical results directly,
calculating in the field of algebraic functions $\overline{\Q(n)}$,
is far from
optimal,  on the other hand
there are known possibilities to avoid some of these complications,
cf.\ for example the methods of partial factorization
and finding dispersion of factors in  \cite{A}, \cite{AB}.

\section*{Acknowledgements} We would like to thank Prof. S.A.~Abramov  and
 Dr. Mark van~Hoeij for the useful information about
their results, Prof. G.P.~Egorychev for constant encouragement and
the Symbolic Computation Group of University of Waterloo for the
support and perfect research environment during my stay there in
2002 when most of the results of this paper were obtained. Remarks of an
anonymous referee (ISSAC'2004) helped to improve the text.

\end{document}